\documentstyle[prabib,aps,epsfig]{revtex}
\begin{document}
\bibliographystyle{prsty}
\newcommand{\dd}{{\rm d}}

\title{Metastable neon collisions: anisotropy and scattering length}
\author{V.P. Mogendorff\footnote{V.P. Mogendorff may be reached at v.p.mogendorff@tue.nl.},
E.J.D. Vredenbregt, B.J. Verhaar and H.C.W. Beijerinck}
\address{Physics Department, Eindhoven University of Technology, P.O. Box 513,\\
        5600 MB Eindhoven, The Netherlands}
\date{\today}
\maketitle

\begin{abstract}

In this paper we investigate the effective scattering length $a$
of spin-polarized Ne$^{*}$. Due to its anisotropic electrostatic
interaction, its scattering length is determined by five
interaction potentials instead of one, even in the spin-polarized
case, a unique property among the Bose condensed species and
candidates. Because the interaction potentials of Ne$^{*}$ are not
known accurately enough to predict the value of the scattering
length, we investigate the behavior of $a$ as a function of the
five phase integrals $\Phi_{\Omega}$ corresponding to the five
interaction potentials. We find that the scattering length has
five resonances instead of only one and cannot be described by a
simple gas-kinetic approach or the DIS approximation. However, the
probability for finding a positive or large value of the
scattering length is not enhanced compared to the single potential
case. We find that the induced dipole-dipole interaction enables
strong coupling between the different $|J\Omega PM_P\rangle$
states, resulting in an inhomogeneous shift of the resonance
positions and widths in the quantum mechanical calculation as
compared to the DIS approach. The dependence of the resonance
positions and widths on the input potentials turns out to be
rather straightforward. The existence of two bosonic isotopes of
Ne$^{*}$ enables us to choose the isotope with the most favorable
scattering length for efficient evaporative cooling towards the
Bose-Einstein Condensation transition, greatly enhancing the
feasibility to reach this transition.

\end{abstract}

\section{Introduction}
\label{sec:intro}

Bose-Einstein Condensation (BEC) has been observed in cold dilute
samples of ground state alkali atoms
\cite{Anderson-1995,Davis-1995,Bradley-1995,Modugno-2001} and
atomic hydrogen \cite{Killian-1998,Fried-1998}. In 2001, the first
condensate of atoms in an electronically excited state was
obtained for metastable He[(1s)(2s)$^3$S$_1$]
\cite{Robert-2001,Dos-Santos-2001}, referred to as He$^{*}$. All
these systems have an electron configuration with only s-electrons
in their open shells in common, resulting in an isotropic
electrostatic interaction.

The other candidate for achieving BEC with atoms in an
electronically excited state is metastable
Ne[(2p)$^{5}$(3s)~$^3$P$_2$], referred to as Ne$^*$ in this paper.
Two groups are pursuing this goal: the group of Ertmer in Hannover
\cite{Zinner-2002} and our group \cite{Kuppens-2002}. Metastable
neon is unique among these species in that its binary
electrostatic interaction is anisotropic, due to its (2p)$^{-1}$
core hole \cite{Doery-1998}.

Crucial in reaching the BEC phase transition is a large ratio of
``good'' to ``bad'' collisions, i.e., a large value of the elastic
collision rate characterized by the total cross section $\sigma =
8\pi a^2 $ for elastic collisions with $a$ the $s$-wave scattering
length, and a small rate for inelastic collisions and other loss
processes. In addition, the creation of a stable BEC requires a
positive value of the scattering length. For metastable rare gas
atoms, such as He$^{*}$ and Ne$^{*}$, the major loss process is
Penning ionization in binary collisions. Fortunately, the latter
process is suppressed in a sample of spin-polarized atoms
\cite{Doery-1998}. For He$^*$, the suppression is very efficient:
only spin flips due to magnetic interactions result in some
residual ionization. Theoretical predictions and recent
experimental data on residual ionization are in good agreement
\cite{Shlyapnikov-1994,Fedichev-1996,{Venturi-1999},{Venturi-2000}}.

For Ne$^{*}$, the anisotropy in the electrostatic interaction
determines the magnitude of the residual ionization in a
spin-polarized gas and has a profound influence on the value of
the scattering length. Theoretical estimates of the rate constant
$\beta^{pol}$ for residual ionization of Ne$^*$ predict a
suppression of ionization by a factor in the range of $10-1000$
\cite{Doery-1998,Derevianko-2000}, depending on the details of the
interaction potentials. In experiments in Hannover and Eindhoven,
a lower limit on the suppression of ionization by a factor of $10$
has been confirmed, but so far no conclusive experimental data on
the residual ionization rate of Ne$^*$ are available.

\begin{figure}
\centerline{\epsfig{figure=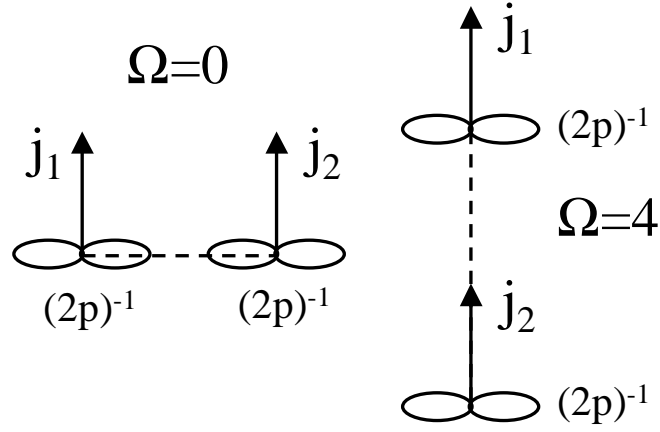,width=100mm}}
\caption{Schematic view of two colliding Ne$^{*}$ $|^3$P$_2\rangle
$ atoms in a spin-polarized  gas ($S=2,J=4$) for both the
$\Omega=0$ and the $\Omega=4$ state. The orientation of the
electronic angular momentum $j_{1,2}$ and the $(2p)^{-1}$ core
hole of the individual atoms is indicated schematically. }
\label{fig:collisions}
\end{figure}

In addition, the anisotropy in the interaction results in
different interaction potentials $V_{\Omega }$ for the molecular
states $|$J, $\Omega \rangle $ of the colliding Ne$^{*}$ atoms,
with $\Omega$ the absolute value of the projection of the total
electronic angular momentum $\vec{J}=\vec{j_1} +\vec{j_2}$ of the
two colliding atoms on the internuclear axis. For binary
collisions of spin-polarized Ne$^{*}$, we have J=4 and $\Omega =
0$ through 4, depending on the relative orientation of the atoms
during the collision. This is illustrated in
Fig.\ref{fig:collisions}, which shows two colliding atoms in the
$\Omega=0$ and $\Omega=4$ state, respectively, with the electronic
angular momentum $j_{1,2}$ and the (2p)$^{-1}$ orbital of the core
hole indicated schematically.

Because the scattering length is determined by the phase integral
of the interaction potential, these different potentials
$V_{\Omega}$ for Ne$^*$ correspond to different scattering lengths
$a_{\Omega }$. Since there is no preference for a certain relative
orientation of the atoms (or $\Omega $ state) during the
collision, even for spin-polarized Ne$^{*}$, the elastic collision
cross section will be determined by an effective overall
scattering length $a$, incorporating the behavior of all five
$\Omega$ states involved. Among the species where Bose-Einstein
condensation has been achieved, the BEC candidate Ne$^*$ thus has
a unique property.

In this paper we investigate the relation between the effective
overall scattering length $a$ and the phase integrals
$\Phi_{\Omega }$ of the potentials $V_{\Omega}$. Although the
potentials of Ne$^*$ are unknown at the level of accuracy needed
to predict the value of the scattering length, it is useful to
investigate the behavior of the effective scattering length as a
function of the average value of the phase integral. In the
following, we will refer to the effective overall scattering
length simply as scattering length.

A better understanding of the complex scattering length of Ne$^*$
is crucial in determining the feasibility of achieving BEC with
Ne$^*$. Important questions that need to be answered to determine
the feasibility for achieving BEC are: (1) Is there a larger
probability of encountering positive values of the scattering
length, as compared to the 75\% probability for the single
potential case? (2) What is the probability for finding a
sufficiently large elastic total cross section for efficient
evaporative cooling? (3) How does the availability of two bosonic
isotopes of Ne$^{*}$, $^{20}$Ne$^*$ and $^{22}$Ne$^*$ (with a
natural abundance of $90$ $\%$ and $10$ $\% $, respectively)
influence these chances?

\begin{figure}
\centerline{\epsfig{figure=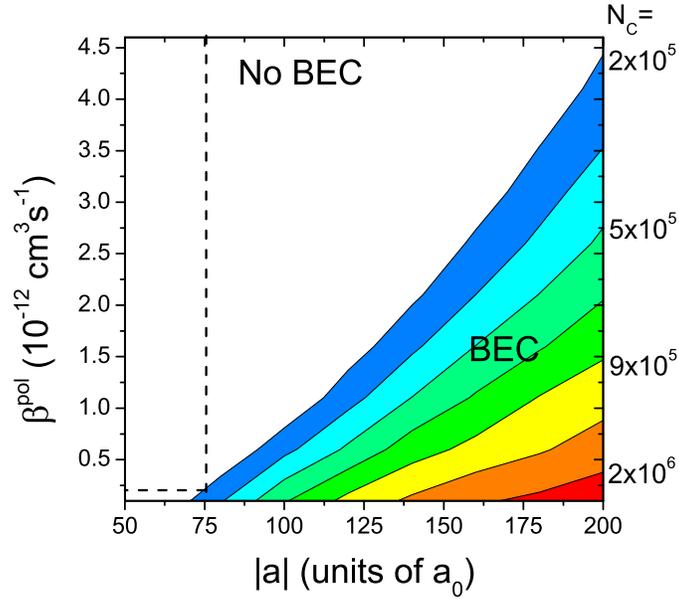,width=100mm}}
\caption{Feasibility plot for reaching the BEC transition with
$^{20}$Ne$^{*}$ in the Eindhoven experiment, showing the number of
atoms $N_c$ with which quantum degeneracy is achieved as a
function of the rate constant $\beta^{pol}$ for residual
ionization and the absolute value of the scattering length $|a|$.
The broken lines correspond to $a_c$ and $\beta^{pol}_c$ for which
BEC of Ne$^{*}$ becomes feasible.}
 \label{fig:feasibilityBEC}
 \end{figure}

The values of the rate constant for residual ionization
$\beta^{pol}$ and the scattering length $a$ which we need to
achieve BEC with Ne$^{*}$ for typical experimental conditions in
our experiment, are summarized in Fig. \ref{fig:feasibilityBEC}.
This so-called feasibility plot for achieving BEC with Ne$^{*}$
shows the calculated number of atoms $N_c$ with which quantum
degeneracy is achieved as a function of $\beta^{pol}$ and the
absolute value of the scattering length $|a|$. The calculation of
this feasibility plot is based on the kinetic model of Luiten {\it
et al} \cite{Luiten-1996}, including trap losses. The criterion
for crossing the BEC transition has been set to $N_c
> 1\times 10^5$ atoms. The initial conditions of the evaporative
cooling process are taken equal to experimental values which we
are able to produce, i.e., $N=1.5\times 10^9 $, $T=1.2$ mK, $\tau
=8 $ s and $\eta = 5.5$, with $N$ the number of atoms, $T$ the
temperature, $\tau$ the lifetime of the atom cloud,
$\eta=E/(k_BT)$ the truncation parameter, $E$ the energy and $k_B$
the Boltzmann constant.

From Fig. \ref{fig:feasibilityBEC} it is clear that BEC of
Ne$^{*}$ is feasible for $\beta^{pol}_c \leq 2.5\times 10^{-13}$
cm$^3$s$^{-1}$ (a suppression of $500$) and $a_c \geq 75$ a$_0$,
with a$_0$ the Bohr radius. Therefore, we calculate the
probability $P_c$ that either of the bosonic isotopes of Ne$^{*}$
has a total cross section larger than $\sigma_c = 8\pi
a_c^2=1.4\times 10^5$ a$_0^2$. A larger suppression of ionization
of course allows for a smaller value of the scattering length and
vice versa.

As a first order estimate, in a simple gas-kinetic picture, one
might expect the effective elastic total cross section $\sigma$,
referred to as elastic cross section in the remainder of this
paper, to be a weighted average over the elastic cross sections
$\sigma _{\Omega } = 8 \pi a_{\Omega}^2$ of the different
molecular states. Applying this approach to the anisotropic
Ne$^{*}$ problem results in a large enhancement of the probability
of encountering large values of $\sigma$ as compared to all
systems with an isotropic interaction potential. Moreover, the
elastic cross section is always larger than a rather large lower
limit.

A more sophisticated approach can be found in the Degenerate
Internal State (DIS) method \cite{Vogels-1998}. In this method,
the energy splitting of the internal states is neglected and $a$
is given by the weighted average of the contributing scattering
lengths $a_{\Omega}$. For cold collisions of hydrogen atoms, the
DIS method results in values for the scattering length that
compare well to the outcome of full quantum calculations
\cite{Verhaar-1987,Stoof-1988}.

The definitive approach to determine the scattering length $a$ of
Ne$^{*}$, of course, is a full five-channel quantum-mechanical
calculation. Although this numerical approach supplies the correct
answer to our problem, it has the disadvantage that the results
are not always easy to understand in terms of the properties of
the input potentials. We use the results of the numerical
calculation to check the validity of the different analytical
approximations described above, which in general give more
insight.

This paper is organized as follows: first, the available ab-initio
potentials and the calculation of the different phase integrals
$\Phi _{\Omega }$ are discussed in Sec. \ref{sec:intpot}. In Sec.
\ref{sec:semiclassical} the single potential scattering length
(Sec. \ref{sec:isoa}), the gas-kinetic approach to defining an
elastic cross section $\sigma$ (Sec. \ref{sec:incoh}) and the
scattering length obtained with the DIS approximation (Sec.
\ref{sec:weighta}) are discussed. We then present the results of
our quantum mechanical, numerical scattering calculation in Sec.
\ref{sec:num}. To end we present our conclusions in Sec.
\ref{sec:concl}. All calculations are performed for both bosonic
isotopes of Ne$^{*}$, $^{20}$Ne$^{*}$ and $^{22}$Ne$^{*}$.

\section{Interaction potentials}
\label{sec:intpot}

There are five adiabatic molecular states $|$J, $\Omega\rangle$
that connect to the spin-polarized Ne$^{*}$ + Ne$^{*}$ asymptotic
limit with total electronic angular momentum $J=4$ and total spin
$S=2$. The degeneracy of the $\Omega =0$ state is 1, that of all
others is equal to 2. As input potentials for Ne$^*$ we have used
the short-range ab-initio potentials of Kotochigova {\sl et al.}
\cite{Kotochigova,Kotochigova-2000}, which are available in the
range $R \leq R_1$, with $R_1 = 60$ a$_0$ for $\Omega = 4$ and
$R_1 = 120$ a$_0$ for all other potentials. Typical values of the
well depth $\epsilon $ and its position $R_m$ are $\epsilon
\approx 30$ meV and $R_m \approx 10$ a$_0$. The long-range
behavior of the potential curves is dominated by the attractive
induced dipole-dipole interaction $-C_6 /R^6 $. The ab-initio
potentials have within $\leq 3$~$\% $ identical $C_6 $
coefficients, since the long-range interaction is dominated by the
(3s) valence electron \cite{Doery-1998,Derevianko-2000}. We use a
single $C_6$ coefficient with a value of $C_6 = 1938$ a.u. as
calculated by Derevianko {\sl et al} \cite{Derevianko-2000}, based
on the static polarizability of neon.

We characterize the potentials $V_{\Omega}$ by their classical
phase integral
\begin{eqnarray}
\Phi_{\Omega}&=&\int_{R_c}^{R_s} k_{\Omega}(R) dR +
                    \int_{R_s}^{\infty} k_{\Omega}(R) dR\nonumber\\
             &=& \Phi_{\Omega}^{R<R_s} + \Phi^{R>R_s},
\label{eq:phaseint}
\end{eqnarray}
with $k_{\Omega}(R)$ the local wave number and $R_c$ the classical
inner turning point for zero collision energy. We choose $R_s$
such that for $R < R_s$ the energy splitting $\Delta
V_{\Omega,\Omega^{\prime}} (R)=V_{\Omega}(R)-V_{\Omega'}(R)$
between the $\Omega$-potentials dominates over the rotational
coupling $\Delta V_{rot} (R)=-[\hbar^2/(2\mu
R^2)]\delta_{|\Omega-\Omega'|=1}\sqrt{(P(P+1)-\Omega\Omega')}
\sqrt{(J(J+1)-\Omega\Omega')}$, with $\mu$ the reduced mass, $l$
the rotational angular momentum and $\vec{P}=\vec{J}+\vec{l}$ the
total angular momentum. For $R > R_s$ the opposite holds.

The first part $\Phi_{\Omega}^{R<R_s}$ of the phase integral has
been calculated by numerical integration in the interval $[R_c,
R_s]$, with $R_s = 20$ a$_0$. The contribution for ${R>R_s}$ to
the phase integral is calculated analytically assuming a pure
long-range $-C_6/R^6$ behavior, resulting in $\Phi^{R>R_s} = 3.34$
$\pi$.

The numerical results are given in Table \ref{tab:phaseintegral}
for both bosonic isotopes $^{20}$Ne$^{*}$ and $^{22}$Ne$^{*}$. We
see that the different $\Omega$-states have very different values
of $\Phi_{\Omega}$, varying by as much as $\Delta\Phi_{\Omega
,\Omega'} = \Phi_{\Omega} - \Phi_{\Omega'} = 0.8~\pi$. This
implies both a different number of bound states and different
positions of the resonances in $a_{\Omega}$. Asymptotic behavior
of, or a resonance in the scattering length occurs when a
quasi-bound state lies close to the dissociation limit or has just
moved into the continuum.

Because the interaction potentials of neon are not known
accurately enough to predict $a$, we have to vary $\Phi_{\Omega}$
over a range equal to $\pi$ to predict the range of $a$ values
that we can expect for the spin-polarized Ne$^{*}$ system.
Therefore, we introduce a scanning parameter $\phi$ that we add to
$\Phi_{\Omega}$ to create a modified phase integral
$\phi_{\Omega}$ according to
\begin{eqnarray}
\phi_{\Omega} &=& \Phi_{\Omega} + \phi, \label{eq:phiscan}
\end{eqnarray}
with $\phi\in [0,\pi]$.

For simplicity, we assume for now that the phase differences
between the $\Omega$ potentials $\Delta \Phi_{\Omega ,\Omega'}$
are constant and equal to the ab-initio values given in Table
\ref{tab:phaseintegral}. Later on in Sec. \ref{sec:num}, we will
also vary $\Delta \Phi_{\Omega ,\Omega'}$ over an interval $\pi$
to investigate the influence of $\Delta \Phi_{\Omega ,\Omega'}$ on
the scattering length $a$.

The classical phase integrals of the two bosonic isotopes of
Ne$^{*}$ are related by the mass scaling rule
\begin{eqnarray}
^{22}\phi_{\Omega} &=& (22/20)^{1/2}~~^{20}\phi_{\Omega}.
 \label{mass_scaling}
\end{eqnarray}
Using the average phase integral over all $\Omega$ states
$\langle\Phi\rangle_{\Omega} = 16.5~\pi $, we find an isotope
shift equal to $0.81~\pi$. This simple relation between the phase
integrals of $^{20}$Ne$^{*}$ and $^{22}$Ne$^{*}$ enables us to
compare very easily the results obtained for $^{20}$Ne$^{*}$ with
those for $^{22}$Ne$^{*}$.

\section{Analytical approach}
\label{sec:semiclassical}

\subsection{Single potential scattering length} \label{sec:isoa}

The semiclassical analysis of the scattering length in atomic
collisions by Gribakin and Flambaum \cite{Gribakin-1993} yields
for the $s$-wave scattering length of a potential with a
long-range behavior $-C_6/R^6$
\begin{eqnarray}
a_{\Omega }&=& a_{bg}[ 1 - \tan ({\phi }_{\Omega } - \frac{\pi}{8})] \nonumber\\
a_{bg} &=& \cos(\pi/4)~[\sqrt{2 \mu
C_6}/4\hbar]^{1/2}~[\Gamma(3/4)/\Gamma(5/4)] ,
 \label{eq:flambaum}
\end{eqnarray}
with $\Gamma ()$ the Gamma function and $a_{bg}$ the background
value of the scattering length. The latter is fully determined by
the long-range behavior of the potential and is equal to $a_{bg}=
44.3$ a$_0$. The position of the resonance in $a_{\Omega }$ is
equal to $\phi_{\Omega}^{res} \bmod \pi = \pi/2 + \pi/8 = 5\pi/8$.
We define the width $\Gamma_{\Omega} $ of a resonance in
$a_{\Omega}$ as
\begin{equation}
\Gamma_{\Omega} =\phi_{\Omega} (3a_{bg})-\phi_{\Omega} (-a_{bg}),
\label{eq:width}
\end{equation}
around the resonance position $\phi^{res}_{\Omega }$. The
scattering length varies around $a_{bg}$ with a probability of
$75$~$\%$ of being positive. The probability that at least one of
the bosonic isotopes of Ne$^{*}$ is positive is much larger: $94$
$\%$.

\subsection{Gas-kinetic model}
\label{sec:incoh}

In a simple gas-kinetic approach, we define the elastic cross
section as a weighted average of the elastic cross sections
$\sigma _{\Omega }$
\begin{eqnarray}
\sigma &=&\langle 8 \pi a_{\Omega}^2\rangle\nonumber\\
    &=&\sigma_{bg}~\sum_{\Omega = 0}^{J} w_{\Omega}^2~
    [1 - \tan{(\phi_{\Omega} - \frac{\pi}{8})}]^2. \label{eqn:sigma}
\end{eqnarray}
Here $\sigma_{bg} = 8\pi a_{bg}^2 = 0.5\times 10^5$ a$_0^2$ is the
background value of the elastic cross section and $w_{\Omega}$ is
the amplitude of the projection of the initial asymptotic
rotational state on the $\Omega$ basis, with $w_0 = 1/3$ and
$w_{\Omega>0} = \sqrt{2}/3$. We assume that we are in the low
temperature limit where the condition $k a \ll 1$ holds, with
$k=\sqrt{2\mu E/\hbar}$ the asymptotic wave number and $E$ the
collision energy in the reduced system.

\begin{figure}
\centerline{\epsfig{figure=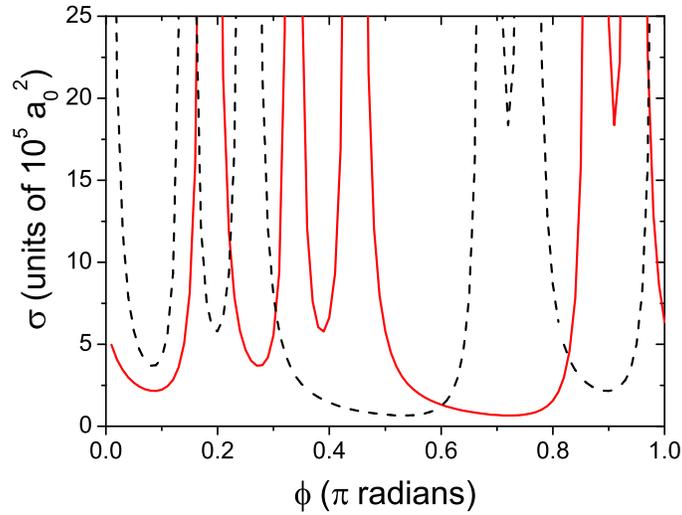,width=100mm}}
\caption{Elastic cross section $\sigma = \langle 8 \pi
a_{\Omega}^2\rangle$ in a gas-kinetic approach for $^{20}$Ne$^{*}$
(solid line) and $^{22}$Ne$^{*}$ (broken line), as a function of
the phase integral $\phi \in [0,\pi]$. We observe five resonances
due to the five contributing potentials, which also results in a
minimum value $\sigma_{min}> 0$ for $\sigma$. }
\label{fig:sigmaomega}
\end{figure}

In Fig. \ref{fig:sigmaomega} we show the results for the elastic
cross section of $^{20}$Ne$^*$ (solid lines) and $^{22}$Ne$^*$
(broken lines) as a function of $\phi \in [0,\pi]$ (Eq.
(\ref{eq:phiscan})). Two important characteristics in the elastic
cross section of Ne$^{*}$ immediately catch the eye. First, we see
a rather large minimum value $\sigma_{min}$ for the elastic cross
section and an increase of almost a factor of two in the
probability $P_c$ that the elastic cross section is large enough
to make BEC of Ne$^{*}$ feasible as compared to the single
potential case, as can be seen in Table \ref{tab:semiclassical}.

Secondly, we see five resonances in $\sigma$, attributable to the
five different $\Omega $-states of Ne$^{*}$. This characteristic
behavior does not depend very much on the specific values of
$\Delta\Phi_{\Omega ,0}$ as long as they are not very small ($\leq
0.05$ $\pi$). This behavior is very different from the single
potential case, where we can encounter an elastic cross section
equal to zero and there exists only one resonance.

The general picture is the same for both isotopes. However, the
elastic cross section of $^{22}$Ne$^{*}$ is shifted with respect
to the elastic cross section of $^{20}$Ne$^{*}$ by the isotope
shift of $0.81$ $\pi $. Depending on the actual phase integral
$\phi_{\Omega}$ of the $^{20}$Ne$^{*}$ system, it can thus be
advantageous to switch to the less abundant bosonic isotope
$^{22}$Ne$^{*}$, to optimize the value of the elastic cross
section. Choosing for each value of $\phi$ the isotope with the
largest $\sigma$ yields an even larger minimum value of $\sigma $,
as can be seen in Table \ref{tab:semiclassical}. In addition,
$P_c$ increases from $78$ $\%$ in the single isotope case to $99$
$\%$ for either isotope.

\subsection{DIS model} \label{sec:weighta}

Next, we investigate the scattering length in the DIS
approximation, which has proven to be quite insightful for
hydrogen \cite{Verhaar-1987,Stoof-1988}. In this approach, the
energy splitting of the internal states is neglected. For
Ne$^{*}$, this means that the rotational splitting $\Delta
V_{rot}$ between the partial waves is neglected. In the DIS
approximation, the scattering length is given by a weighted
average of the five $a_{\Omega }$'s involved \cite{Stoof-1988}
\begin{eqnarray}
a &=& \langle a_{\Omega} \rangle \nonumber\\
    &=& a_{bg}~~\sum_{\Omega =0}^4 w_{\Omega }[ 1 - \tan (\phi_{\Omega} - \frac{\pi}{8})].
\label{eqn:aweight}
\end{eqnarray}
From Eq. (\ref{eqn:aweight}) it is clear that the resonance
positions $\phi_{\Omega}^{res}$ of $a$ in the DIS approximation,
coincide with the single potential resonance positions. They are
completely determined by the values of $\Delta\Phi_{\Omega,0}$
\begin{eqnarray}
\phi_{\Omega}^{res}&=& (\frac{5}{8}\pi - \Delta \Phi_{\Omega,0})
\bmod\pi . \label{eq:posres}
\end{eqnarray}

\begin{figure}
\centerline{\psfig{figure=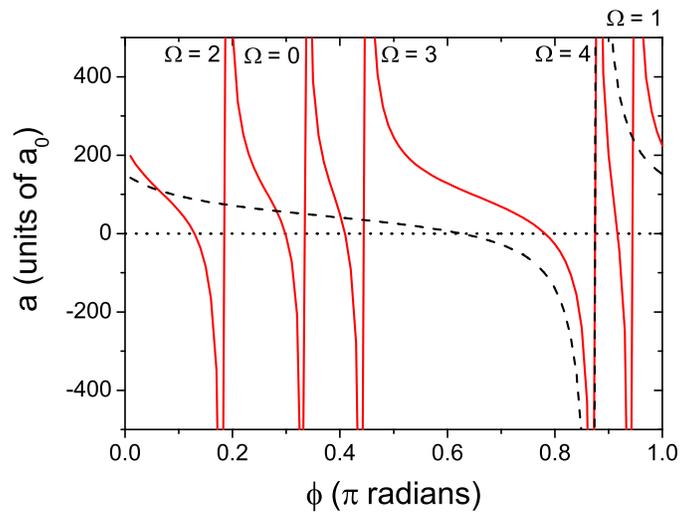,width=100mm}}
\caption{DIS result for the scattering length $a$ (solid line) as
a function of the parameter $\phi \in [0, \pi]$, showing five
resonances that are labelled with an $\Omega$-value because they
are located at the position of the single potential resonances in
$a_{\Omega}$. For comparison we show the behavior of $a_{4}$
(broken line).}
 \label{fig:avalue}
\end{figure}

Figure \ref{fig:avalue} shows the scattering length $a$ as a
function of the scanning parameter $\phi \in [0, \pi]$. Again we
observe five resonances in the scattering length. In this figure
we have also plotted the behavior of the single potential
scattering length $a_{\Omega =4}$ (broken line), showing clearly
that the single potential resonance positions coincide with the
resonance positions of $a$ in the DIS approach.

Taking the weighted average of $a_{\Omega }$ does not change the
total probability for a positive scattering length ($75$ $\%$) as
compared to the single potential case (Sec. \ref{sec:isoa}).
However, the probability of encountering a large value of $a$ does
increase, as can be seen in Table \ref{tab:semiclassical}. Both
the width of the resonances (Eq. (\ref{eq:width})) and the
derivative $\partial a/\partial\phi$ of $a$ with respect to $\phi$
determine this probability. From Fig. \ref{fig:avalue} it is clear
that the total width $\Gamma$ of all five resonances combined is
much larger than the single potential resonance width. The width
of each resonance is not only determined by the relative weight
$w_{\Omega }$ of its single potential scattering length but also
by its relative position $\phi^{res}_{\Omega}$ in respect to the
other resonances, and therefore depends sensitively on the phase
differences $\Delta\Phi_{\Omega,\Omega'}$ between the potentials.
An increased width and derivative of $a$ in the DIS approach as
compared to the single potential case therefore lead to a much
larger $P_c$ (Table \ref{tab:semiclassical}). Choosing the most
advantageous bosonic isotope improves again $\sigma_{min}$ and
$P_c$ as well as the probability for encountering a positive value
of $a$ ($95$ $\%$) as compared to the single isotope case (Table
\ref{tab:semiclassical}).

\section{Quantum mechanical numerical calculation}
\label{sec:num}

\subsection{Model}
\label{ssec:model}

\begin{figure}
\centerline{\psfig{figure=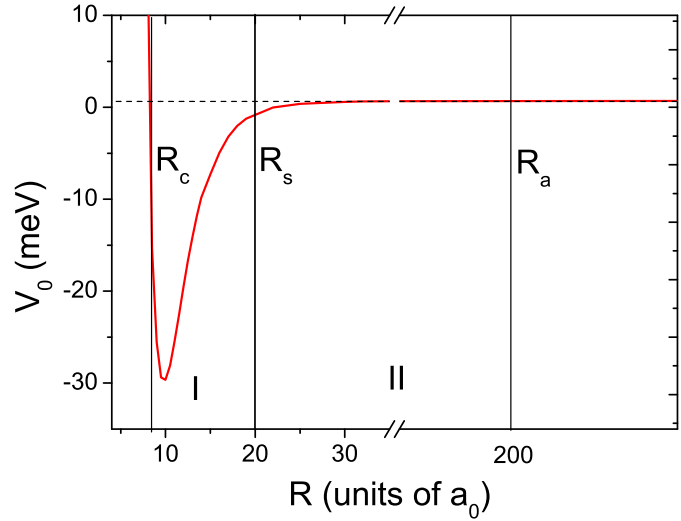,width=100mm}}
\caption{Interaction potential $V_{0}$ for the $\Omega=0$
molecular state of two colliding spin-polarized Ne$^{*}$ atoms,
with both range I (dominant $\Omega$-splitting) for $R \leq R_s =
20$ a$_0$ and range II (dominant rotational energy splitting) for
$R_s < R < R_a = 200$ a$_0$ indicated. }
 \label{fig:intregion}
 \end{figure}

For the Ne$^{*}$ system, we can distinguish two regions of
interest in the potentials $V_{\Omega}(R)$  (Fig.
\ref{fig:intregion}). At small internuclear separations (region
I), i.e. $R<R_s$, the splitting $\Delta V_{\Omega ,\Omega'}(R)$
between the interaction potentials $V_{\Omega}$ is larger than the
differential rotational coupling $\Delta V_{rot}(R)$. As a result,
there is no coupling between different $\Omega $ states but
coupling between different $l$ states. In this region $\Omega $ is
a good quantum number and the $|J\Omega P M_P\rangle $ basis is
the proper representation. Here $M_P$ is the projection of $P$ on
the quantization axis. Both $P$ and $M_P$ are conserved during the
collision. In a semiclassical picture we can describe this regime
in the following way. The large $\Omega$-splitting results in a
rapid precession of $\vec{\rm J}$ about the internuclear axis,
much faster than the rotation of the axis itself as determined by
the value of $|\vec{l}|$. The projection of $\vec{\rm J}$ on the
internuclear axis is thus conserved, while the magnitude of
$\vec{l}$ changes due the changing orientation of $\vec{\rm J}$
with respect to the space-fixed total angular momentum vector
$\vec{P}$.

At large internuclear separations (region II in Fig.
\ref{fig:intregion}), i.e. $R>R_s$, the rotational coupling
dominates the interaction: the splitting $\Delta V_{\Omega
,\Omega'}(R)$ is smaller than the differential rotational coupling
$V_{rot}(R)$. The relative motion of the colliding atoms results
in a rotation of the internuclear axis with respect to $\vec{\rm
J}$ and therefore in a change in the value of $\Omega $
(Fig.\ref{fig:collisions}). In this region $\Omega $ is not a good
quantum number but $l$ is, and we use the $|JlPM_P\rangle $
representation.

Ultra-cold collisions between spin-polarized Ne$^{*}$ atoms are
described by a five-channel problem: five $\Omega $= $0..4$
channels in region I and five $l=0,2,..8$ channels in region II.
Only even partial waves contribute due to the symmetry requirement
of the wave function for bosons. The rotational energy barrier ($
5.6$ mK for l=2, located at $78$ a$_0$) is always much larger than
the collision energy ($\leq 0.5$ mK). For this reason, higher
order partial waves ($l \neq 0$) do not contribute to the incoming
channel. However, in region I, the short-range interaction $\Delta
V_{\Omega, \Omega'}$ still couples the single incoming channel
$|J=4$ $l=0$ $P=4$ $M_P=4\rangle$ to higher order partial waves.
Because the tunnelling probability for $l \neq 0$ is negligible
($\leq 10^{-5}$), they only contribute to the elastic scattering
process when they again couple to the $|J=4$ $l=0$ $P=4$ $M_P
=4\rangle$ initial state.

In our calculations we assume that the intermediate region, where
$V_{rot}(R)$ and $V_{\Omega,\Omega'}(R)$ are of the same order of
magnitude, is arbitrarily small, i.e., we assume a sudden
transition from region I to region II at $R = R_s$. The scattering
problem now reduces to potential scattering and we can solve the
uncoupled problem in region I in the $|J\Omega P M_P\rangle$-basis
and in region II in the $|JlP M_P\rangle$-basis, for each channel.
After transformation of the solution $u_I(R)$ in region I at
$R=R_s$ from the $|J\Omega P M_P\rangle$-basis to the $|JlP
M_P\rangle$-basis, we connect it continuously to the long-range
solution $u_{II}(R)$ at $R_s$, assuming equal local wave numbers
\cite{Vogels-1998}.

In very good approximation \cite{Verhaar-1993,Moerdijk-1994}, we
can summarize the behavior of the atoms in region I by means of
the accumulated phase method. In $R=R_s$ the radial wave function
of a single uncoupled channel is then given by
\begin{equation}
{u}_{I}^{\Omega }(R_s) = \sin(\Phi_{\Omega}^{R<R_s} + \phi +
\pi/4), \label{eq:yshort}
\end{equation}
with the extra phase shift $\pi/4$ due to our choice of using the
classical phase rather than the quantum mechanical accumulated
phase. After transformation to the $|JlPM_P\rangle$ basis the
$5\times 5$ solution matrix is given by
\begin{equation}
{u}_{I}^{l}(R_s) = T_{l\Omega} {u}_{I}^{\Omega }(R_s),
\label{eq:yshort2}
\end{equation}
with $T_{l\Omega }$ the elements of the transformation matrix
$\underline{\bf {T}}$ between the $|J \Omega P M_P \rangle$ and
the $|JlPM_P \rangle$ basis.

In region II, the evolution of the radial wave function
$u_{II}^{l}$ for $R>R_s$ is governed by the radial Schr\"odinger
equation
\begin{equation}
\frac{{\partial }^2}{\partial R^2}{u}_{II}^{l}(R) + \left(
k^2-\frac{l(l+1)}{R^2}-
    \frac{2\mu C_6}{{\hbar}^2R^6}\right ) {u}_{II}^{l}(R)=0.
\label{eq:s}
\end{equation}
The connection of the short (I) and long-range (II) solutions in
the $|JlPM_P\rangle$ representation at $R=R_s$ determines the
boundary conditions for the numerical integration of $u_{II}^{l}$
from $R_s$ to $R_a$, where the asymptotic limit of a vanishing
potential is valid. At $R_a$, the numerical solution is connected
to the asymptotic radial wave function
\begin{equation}
u_{II}^{l}(R_a) = \frac{1}{\sqrt{k}}[A_{l} e^{-i(kR_a-l\pi
/2)}+B_{l} e^{i(kR_a-l\pi /2)})], \label{eq:yasymp}
\end{equation}
to determine the scattering matrix $\underline{\bf
S}=\underline{\bf B}~\underline{\bf A}^{-1}$. From the scattering
matrix we then obtain the scattering length $a$
\begin{equation}
a=-\lim_{k\rightarrow 0}\frac{\tan{(\ln{(S_{00})}/2i)}}{k}.
\label{eq:asmatrix}
\end{equation}


\subsection{Results}\label{ssec:numresult}

\begin{figure}
\centerline{\psfig{figure=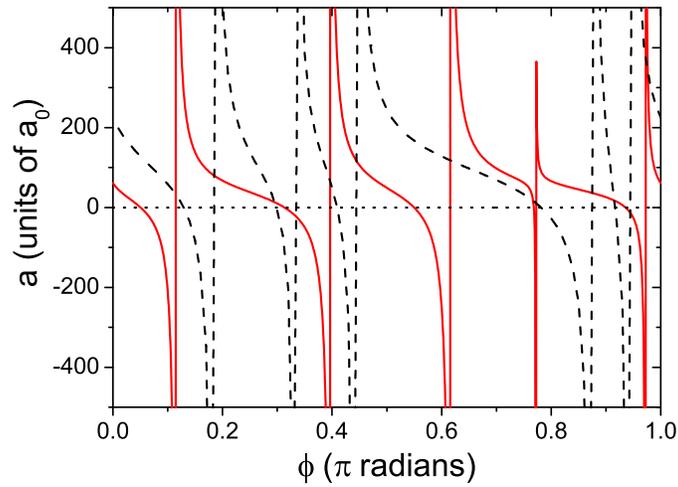,width=100mm}}
\caption{Full quantum mechanical calculation of the scattering
length $a$ (solid line) as a function of $\phi \in [0, \pi]$. For
comparison we also have depicted the scattering length $a$
obtained with the DIS model (broken line). Both position and width
of the resonances in $a$ differ from the DIS result: an
$\Omega$-label cannot be attached to any separate resonance.}
 \label{fig:anumweight}
\end{figure}

Using the ab-initio potentials of section \ref{sec:intpot} we have
calculated the scattering length of $^{20}$Ne$^*$ using the
quantum mechanical calculation described in section
\ref{ssec:model}. Again, we have performed these calculations for
$\phi \in [0, \pi]$, to determine the range of scattering length
and elastic cross section values that we can expect for the
Ne$^{*}$ system. Figure \ref{fig:anumweight} shows the scattering
length $a$ (solid line) as a function of $\phi$. For comparison,
we have included the result for $a$ obtained with the DIS model
(broken line) (Subsec. \ref{sec:weighta}), in which the positions
of the resonances in $a$ correspond to the single potential
resonance positions in $a_{\Omega}$. The resonance positions
$\phi^{res}_{\Omega}$ of $a$ are shifted in an inhomogeneous way
with respect to those obtained with the DIS method and also the
widths $\Gamma_{\Omega}$ of the individual resonances differ
significantly. We can no longer attach an $\Omega$-label to each
of the resonances, as is possible in the DIS approach. In the DIS
model, we neglect the rotational splitting $V_{rot}$ between the
partial waves, so that effectively all partial waves are
equivalent. Apparently, both the resonance positions as well as
the widths of the individual resonances in the numerical results
for $a$ are influenced by the coupling to higher order partial
waves. This is not surprising: close lying bound states from other
$\Omega$-potentials will most likely shift the bound state and
thus the corresponding resonance position. Clearly, neglecting the
rotational splitting of the internal states is not justified in
the case of Ne$^{*}$.

However, despite these qualitative differences in the behavior of
$a$, the quantitative behavior is the same as in the single
potential case. The probabilities for encountering a positive $a$
value and an elastic cross section $\sigma_c$ are equal to those
found in the single potential case (Table
\ref{tab:semiclassical}). Although the total width of the
resonances is larger, this is compensated by a decrease in the
derivative of $a$. Moreover, choosing the most advantageous
bosonic isotope for each value of the scanning parameter $\phi$
leads to a similar increase in these values for both the full
quantum mechanical calculation as the single potential case.

\bigskip

To investigate the influence of $\Delta \Phi_{\Omega ,\Omega'}$ on
the positions of the resonances in $a$, we have varied the
classical phase difference $\Delta\Phi_{4,0}$ of one of the
$\Omega$ potentials ($\Omega=4$), while keeping the other
classical phase differences fixed at their ab-initio value (Table
\ref{tab:phaseintegral}). Starting at its ab-initio value, the
classical phase difference
\begin{equation}
\Delta\phi_{4,0} = \Delta\Phi_{4,0} +\Delta\phi,
\label{eq:phasediff}
\end{equation}
 with
$\Delta\phi\in [0,\pi]$, is varied over $\Delta\phi_{4,0} \in
[-0.54~\pi, 0.46~\pi]$. In this way, the bound states in the
$\Omega=4$ potential encounter the bound states in all other
$\Omega $ potentials. The position of the resonances
$\phi^{res}_{\Omega}$ for this modified set of ab-initio
potentials is determined in the usual way, by scanning the
parameter $\phi$ in the phase integral $\phi_{\Omega}$ (Eq.
(\ref{eq:phiscan})) over the range $[0 ,\pi]$.

\begin{figure}
\centerline{\psfig{figure=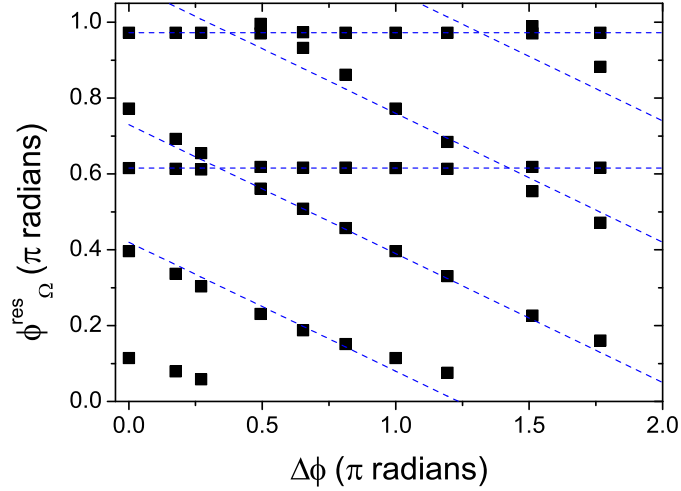,width=100mm}}
\caption{Numerical calculation of the phase integrals
$\phi^{res}_{\Omega}$ (filled squares) at which a resonance in $a$
occurs for a modified set of potentials, with the phase integral
difference $\Delta\phi_{4,0}$ varying from its ab-initio value
$-0.54~\pi$ to $+1.5\pi$ by varying $\Delta\phi$ over $2\pi$. In
this way, the bound levels of the $\Omega=4$ state `encounter' the
bound levels in all other $\Omega$-states. The broken lines are
drawn to guide the eye.}
 \label{fig:resanumj4}
\end{figure}

The results are shown in Fig. \ref{fig:resanumj4}, where we have
plotted the position $\phi^{res}_{\Omega}$ of the five resonances
in $a$ as a function of the shift $\Delta\phi$ in the bound states
of the $\Omega=4$ potential. The broken lines are drawn to guide
the eye. We observe that two of the resonances remain at a fixed
position, while all three others shift proportional to
$\Delta\phi$ with a slope equal to $(-0.34\pm 0.02)\pi$ and
separated by $(0.30\pm 0.02)\pi$. Narrow avoided crossings occur
when 'constant' $\phi^{res}_{\Omega}$ meet $\phi^{res}_{\Omega}$
varying like $\propto -\Delta\phi$ at $\Delta\phi = (0.46+n)\pi $
and $\Delta\phi = (0.52+n)\pi $, with $n=0,1,2,..$. In addition,
very broad avoided crossings occur between $\phi^{res}_{\Omega}$
varying like $\propto -\Delta\phi$ at $\Delta\phi\approx (0.1 +
n)\pi$, with $n=0,1,2,..$. We have plotted $\phi^{res}_{\Omega}$
over a range $\Delta\phi \in [0,2\pi]$, to show the avoided
crossings. Varying one of the other classical phase differences
$\Delta\phi_{\Omega\neq 4,0}$ yields similar results.

Clearly, the quasi-bound states of the system are no longer pure
$|J\Omega PM_P\rangle$ states. Apparently, the quasi-bound states
whose resonance positions $\phi^{res}_{\Omega}$ vary proportional
to $\Delta\phi$ are strongly coupled to the $|J\Omega =4
PM_P\rangle$ state and those whose $\phi^{res}_{\Omega}$ remain
constant are only very weakly coupled to the $|J\Omega =4
PM_P\rangle$ state. When two strongly coupled resonances approach
each other, a strong, broad avoided crossing occurs. Similarly, a
weak, narrow avoided crossing occurs when a weakly coupled
resonance approaches one of the other resonances.

\begin{figure}
\centerline{\psfig{figure=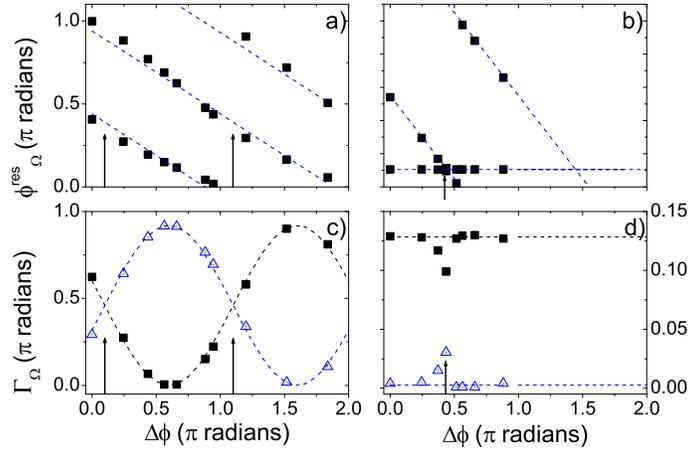,width=100mm}}
\caption{Numerical two-channel ($J=1$) calculation of position (a
and b) and width (c and d) of the resonances in $a$ for a modified
set of potentials, with $\Delta\phi_{1,0}$ varying from its
ab-initio value $-0.61$ $\pi$ to $+1.39$ $\pi$ by varying
$\Delta\phi$ over $2\pi$. The figures on the right hand side (b
and d) show the results for a reduced $C_6^{\prime}=C_6/100$. The
broken lines are drawn to guide the eye and the arrows indicate
the avoided crossings. In the presence of an induced dipole-dipole
interaction (a) the resonance positions vary both proportional to
$\Delta\phi$ and are strongly coupled, and the width of the
resonances (c) varies as a sine. For a small dipole-dipole
interaction (b) one of the resonance positions remains constant,
while the other varies proportional to $\Delta\phi$. The width of
the resonances (d) remains constant, except at the avoided
crossing where they coincide (arrow).} \label{fig:combi}
\end{figure}

This picture is consistent with the behavior of the width of the
resonances as a function of $\Delta\phi$. This becomes clear when
we look at a simpler, more transparent two-channel, i.e. $J=1$,
numerical calculation of the scattering length. The results of
this calculation are consistent with the full five-channel
calculation. In Fig. \ref{fig:combi}a) we have plotted the phase
integrals $\phi^{res}_{\Omega}$ at which a resonance in $a$ occurs
as a function of $\Delta\phi$ for $J=1$. Both resonance positions
vary proportional to $\Delta\phi$ with a slope equal to $(-0.50\pm
0.02)\pi$, and with broad avoided crossing between them at
$\Delta\phi = (0.1 + n)\pi$, with $n=0,1,..$. They are strongly
coupled, which is also reflected in the behavior of the width of
the resonances $\Gamma_{\Omega }$ (Figure \ref{fig:combi}c)),
which varies as a sine (broken line) between $0$ and $\Gamma$. At
the avoided crossings the widths of the resonances become equal.
The total width of both resonances combined $\Gamma = \Gamma_0 +
\Gamma_1 $ is conserved, but one resonance is wide while the other
is narrow. This periodic change in the resonance width is due to
the periodic change in the coupling of both quasi-bound states to
the incoming $l=0$ channel with changing $\phi^{res}_{\Omega}$.

The weakly coupled case can be illustrated by reducing the value
of $C_6$, because in the absence of an induced dipole-dipole
interaction $-C_6/R^6$ no coupling between the quasi-bound states
is possible. This can be understood in the following way. With a
decreasing dipole-dipole interaction the rotational energy barrier
increases both in height and width. As a result, the region around
$R_s$ in which rotational coupling between the different $\Omega$
states takes place decreases and eventually vanishes. Figure
\ref{fig:combi} b) and d) show the resonance positions and widths
for a vanishing value of $C_6$, simulated by assuming a modified
value $C_6' = C_6/100$. One resonance position remains constant,
while the other varies proportional to $\Delta\phi$ and the widths
of the resonances are small and remain constant, except at the
avoided crossing where they approach each other. This is the
behavior seen for some of the resonances in $a$ for $J=4$ and is
qualitatively the same as the behavior of the resonance positions
and widths in the DIS model, where coupling between the
quasi-bound states is not taken into account.

\section{Concluding remarks}
\label{sec:concl}

Elastic collisions between spin-polarized Ne$^{*}$ atoms are
governed by multiple interaction potentials. This unique property
of Ne$^{*}$ among the BEC species and candidates is a result of
the anisotropic interaction between them. Both simple analytical
and full quantum mechanical calculations of the scattering length
$a$ of Ne$^{*}$ show that the resulting scattering length has five
resonances. A simple gas-kinetic picture yields very favorable but
unrealistic results for the elastic collision cross section that
are not compatible with the numerical calculations. This approach
is only valid for an incoherent mixture of $\Omega$ states.
Comparison between the numerical results and the DIS model reveals
that $a$ is also not simply a weighted average over the single
potential resonances $a_{\Omega}$ and that the resonances in $a$
cannot be assigned to a single $\Omega$ state. Although the DIS
approach assumes a coherent mixture of $\Omega$ states, coupling
between the quasi-bound states is not taken into account, and it
therefore does not describe the Ne$^{*}$ system accurately. The
overall behavior of $a$ is similar to that of the usual single
potential scattering length: neither the probability for
encountering a positive or a large value of $a$ is enhanced by the
presence of five instead of one resonances.

The presence of an induced dipole-dipole interaction leads to
strong coupling between the different $\Omega$ states and causes a
broadening of the resonances, resulting in quasi-bound states that
are a linear combination of different $|J\Omega PM_P\rangle$
states. This coupling between the different $|J\Omega PM_P\rangle$
states in turn leads to the inhomogeneous shift of the resonance
positions and widths in the quantum mechanical calculation as
compared to the DIS approach. However, the dependence of the
resonance positions and widths on the input potentials is quite
straightforward. The resonance positions vary either directly
proportional to the relative phase differences between the
$\Omega$ potentials or not at all, depending on the exact
composition of its quasi-bound state. The width of the strongly
coupled resonances (whose positions vary $\propto -\Delta\phi$) is
determined by the coupling to the ingoing channel, which varies
periodically with $\Delta\phi$. The width of the weakly coupled
resonances (whose positions do not vary with $\Delta\phi$) is
constant, since the coupling to the ingoing channel does not
change. The total change in all five resonance positions is always
equal to $-\Delta\phi$ and the total width is conserved.

The possibility to choose between the two bosonic isotopes of
Ne$^{*}$ to optimize the value of the elastic cross section,
greatly enhances the prospects for achieving BEC with Ne$^{*}$
(Table \ref{tab:semiclassical}). Large beam fluxes of both bosonic
isotopes, crucial in obtaining favorable initial conditions for
efficient evaporative cooling, have been realized at the Eindhoven
experiment \cite{Tempelaars-2002}, therefore choosing the isotope
with the most favorable scattering length is feasible.

\section*{Acknowledgements}
We would like to thank S. Kotochigova for calculating the
ab-initio potentials of spin-polarized Ne$^{*}$ and S. Kokkelmans
for careful reading of the document. This work was financially
supported by the Netherlands Foundation on Fundamental Research
and Matter (FOM).



\newpage

\newpage
\begin{table}
 \caption{Classical phase integral $\Phi_{\Omega}$ and its difference $\Delta \Phi_{\Omega,0} =
 \Phi_{\Omega}  - \Phi_0$  with respect to the $\Omega=0$ potential of the spin-polarized
 adiabatic molecular $\Omega$-states, connecting to the Ne$^{*}$ + Ne$^{*}$
 asymptotic limit with J$=4$ and S$=2$. The $\Omega$ states are labelled by $\Omega_g$,
 where the gerade label $g$ reflects the symmetry of the electron
 wave function under inversion around the center of charge.
 Data are given for both bosonic isotopes $^{20}$Ne$^{*}$ and $^{22}$Ne$^{*}$ of Ne$^{*}$.}
 \begin{tabular}{lcccc}
          & \multicolumn{2}{c}{$^{20}$Ne$^{*}$}
          & \multicolumn{2}{c}{$^{22}$Ne$^{*}$}\\
 \hline
  $\Omega$ & $\Phi_{\Omega}$($\pi $ radians)    & $\Delta\Phi_{\Omega,0}$ ($\pi$ radians) & $\Phi_{\Omega}$ ($\pi $ radians)   & $\Delta\Phi_{\Omega,0}$ ($\pi$ radians) \\
 \hline
 $4_g $ &  16.43 & -0.54 & 17.23 & -0.77   \\
 $3_g $ &  16.86 & -0.11 & 17.68 & -0.34   \\
 $2_g $ &  16.12 & -0.85 & 16.91 & -1.09   \\
 $1_g $ &  16.36 & -0.61 & 17.16 & -0.84   \\
 $0_g $ &  16.97 & 0     & 18.00 & 0       \\
 \end{tabular}
 \label{tab:phaseintegral}
 \end{table}

\begin{table}
\caption{Minimum value $\sigma_{min}$ of the elastic cross section
and probability $P_c$ for $\sigma\geq\sigma_{c}=1.4\times 10^5$
$a_0^2$ for a single isotope ($^{20}$Ne$^{*}$) and the set of two
bosonic isotopes $^{20}$Ne$^{*}$ and $^{22}$Ne$^{*}$, as
calculated with the single potential semiclassical model, the
gas-kinetic model, the DIS model and the quantum mechanical
numerical calculation.}
\begin{tabular}{lcccc}
& \multicolumn{2}{c}{single isotope} &\multicolumn{2}{c}{either isotope} \\
 model  & $\sigma_{min}$ (units of $10^5$ a$_0^2$) & $P_c$ ($\%$) & $\sigma_{min}$ (units of $10^5$ a$_0^2$)  &  $P_c$ ($\%$)\\
\hline
single potential &  0 & 42 & 0.16 & 61 \\
gas-kinetic  & 0.7 &  78 & 1.3 & 99 \\
DIS          & 0 & 72 & 0.35 & 95 \\
numerical & 0 & 42 & 0.17 & 67 \\
\end{tabular}
\label{tab:semiclassical}
\end{table}


\end{document}